\documentstyle[preprint,prb,aps]{revtex}

\begin{document}

\tighten
\draft

\title{General relation between state density and dwell times
in mesoscopic systems}
\author{G. Iannaccone}
\address{
Institute for Microstructural Sciences, National Research Council of Canada,
Ottawa, Canada K1A 0R6; \\
Dipartimento di Ingegneria dell'Informazione:
Elettronica, Informatica e Telecomunicazioni \\
Universit\`a degli studi di Pisa, Via Diotisalvi 2,
I-56126 Pisa, Italy}
\date{Published on Phys. Rev. B {\bf 51}, 4727 (1995). -
quant-ph/9511004}
\maketitle

\begin{abstract}
A relevant relation between the dwell time and the density of states
for a three dimensional system of
arbitrary shape with an arbitrary number of incoming channel is derived.
This result extends the one obtained by Gasparian et al.
for the special case of a layered one dimensional
symmetrical system.
We believe that such a strong relation between the most
widely accepted time related to the dynamics of a particle and the density of
states in the barrier region, one of the most relevant
properties of a system in equilibrium, is
rich of physical significance.
\end{abstract}

\pacs{PACS numbers: 03.65.-w, 71.20.-b, 73.20.At, 73.40.Gk}

\section{Introduction}
In the controversial field of tunneling times, the dwell time
is widely accepted as the average time spent by a particle
in a given region of space. This time was postulated
by B\"uttiker in the
context of the tunneling time problem
\cite{but_larmor83}; recently, a rigorous
derivation of the dwell time was obtained within Feynman's
\cite{sokobask87} and Bohm's \cite{leavaers93} formulations of
quantum mechanics.
We should also mention that the agreement on the physical
significance of the dwell time is not unanimous \cite{sokoconn93,olkoreca92}.

The connection between the dwell time and the density of states
in the barrier region was shown by
Gasparian et al.\cite{gasppoll93,gasparian94},
for the case of a one dimensional symmetrical multilayered system.
In this paper we will extend this result to an
arbitrary three dimensional region,
using a very concise derivation.

We believe that establishing and clarifying the connection between these
two quantities is also relevant to the recent work by Wang et al.
\cite{wangguo94}, whose object is the investigation of
the statistics of quasi-bound states in classically chaotic systems.
These authors assume a relation between the energies of quasi-bound
states and the extrema of the dwell time in the system under
consideration, which can be addressed using the results we
present here.

\section{Dwell time and density of states}
Let us consider a region $\Omega$ in three dimensional
space, connected with the outside
world by $N$ channels. Let a given channel be
characterized not only by its spatial location, but also
by its particular propagation mode.

We choose as an orthonormal basis for this system
the set of stationary state  eigenfunctions corresponding
to each of these $N$ incoming channels, i.e. $\{ | \phi_n(E) \rangle \}$,
where $n = 1 \dots N$. Our basis is continuous and degenerate,
so that an appropriate normalization is
\begin{equation}
\langle \phi_n(E') | \phi_m(E'') \rangle
=
\delta_{nm} \delta(E'-E'')
\label{norma}
;\end{equation}
of course, we also have
\begin{equation}
\sum_{n=1}^{N} \int dE | \phi_n(E)\rangle \langle \phi_n(E) | = \hat{1}
\label{unity}
.\end{equation}

We now derive an expression for the dwell time
associated with
a general wave function in this set. Let us consider a wave packet
$| \psi_n(t) \rangle$ incoming from the $n$th channel, such that the
probability of finding the particle in $\Omega$ vanishes for time
approaching $\pm \infty$. $| \psi_n(t) \rangle$ can be written in
terms of our basis as
\begin{equation}
| \psi_n(t) \rangle = \int \alpha_n(E) e^{-iEt/\hbar}
			|\phi_n(E) \rangle dE
\label{spectrum}
,\end{equation}
where $\int |\alpha_n(E)|^2 dE = 1$ so that  $| \psi_n(t) \rangle$ is
normalized to unity.

The mean dwell time in $\Omega$ associated with the wave packet
$| \psi_n(t) \rangle$ is \cite{jawoward_phase88,leavaers_dwell89}
\begin{eqnarray}
\tau_D^{(n)} & = & \int_{-\infty}^{+\infty} dt
		\int_{\Omega} |\psi_n({\bf r},t)|^2 d{\bf r}
			\nonumber \\
	& =&  \int_{-\infty}^{+\infty} dt
		\langle \psi_n(t)|\hat{P}_\Omega|\psi_n(t) \rangle
\label{dwelltime}
,\end{eqnarray}
where $\hat{P}_\Omega$ in the projection operator onto the region $\Omega$.
Substitution of (\ref{spectrum}) in (\ref{dwelltime}) yields
\begin{equation}
\tau_D^{(n)} = 2 \pi \hbar \int
	|\alpha_n(E)|^2 \langle \phi_n(E)|\hat{P}_\Omega|\phi_n(E)\rangle dE
\label{dwellspectrum}
.\end{equation}
Therefore the dwell time $\tau^{(n)}_D(E)$ for the stationary state
$| \phi_n(E) \rangle$ can be defined as
\begin{equation}
\tau^{(n)}_D(E) = 2 \pi \hbar
\langle \phi_n(E)|\hat{P}_\Omega|\phi_n(E)\rangle
\label{dwellstat}
,\end{equation}
which corresponds to the limit of (\ref{dwellspectrum}) as
$|\alpha_n(E)|^2$ tends to a delta function.

Equation (\ref{dwellstat}) is equal to the well-known expression
postulated by
B\"uttiker \cite{but_larmor83}, provided the different normalization
for the wave function used here is taken into account.
In fact, for the wave functions of our basis
the incoming probability current is just $(2\pi \hbar)^{-1}$ (the
total probability is given in units of inverse energy, according
to normalization (\ref{norma}), therefore the probability current
is in units of inverse action).
In the Appendix two interesting formulas relating the dwell
time (\ref{dwellstat}) to perturbative potential
approaches\cite{sokobask87,iannpell94}
and to the Green's functions for our system are shown.

The local density of states $\rho({\bf r},E)$ is given by
\cite{economou79}
\begin{equation}
\rho({\bf r},E) =
\sum_{n=1}^N \int
\langle {\bf r}| \phi_n(E')\rangle
   \langle \phi_n(E')|{\bf r}\rangle
	\delta(E-E') dE'
\label{dos}
,\end{equation}
which, in our case, becomes
\begin{equation}
\rho({\bf r},E) =
\sum_{n=1}^N
   \langle {\bf r}| \phi_n(E)\rangle
   \langle \phi_n(E)|{\bf r}\rangle
\label{dos_inourcase}
.\end{equation}

The density of states $\rho_\Omega(E)$ for
the region $\Omega$ is
just the integral of $\rho({\bf r},E)$ over $\Omega$, therefore we
obtain
\begin{eqnarray}
\rho_\Omega(E) & = & \sum_{n=1}^N
	\int_\Omega
   \langle {\bf r}| \phi_n(E)\rangle
   \langle \phi_n(E)|{\bf r}\rangle
	d{\bf r}
	\nonumber \\
& = & \sum_{n=1}^N
	\langle \phi_n(E)|\hat{P}_\Omega|\phi_n(E)\rangle
\label{dosomega}
.\end{eqnarray}
{}From (\ref{dwellstat}) and (\ref{dosomega}), one
straightforwardly obtains
\begin{equation}
\rho_\Omega(E) =
\frac{1}{2 \pi \hbar}
\sum_{n=1}^N \tau_D^{(n)}(E)
\label{final}
,\end{equation}
i.e., the density of states in $\Omega$ is proportional to the
sum of the dwell times in $\Omega$ for all incoming channels.

\section{Comments}
The result of Gasparian et al.\cite{gasppoll93}
can be easily obtained as a particular case of (\ref{final}).
In fact, for a one dimensional region the number of channels
reduces to two (for left and right incoming waves) and for a
symmetric potential we have $\tau^{(1)}_D(E) =\tau^{(2)}_D(E) = \tau_D(E)$,
so that we can write
\begin{equation}
\rho_{\mbox{[1dim-sym]}} = \frac{1}{\pi \hbar} \tau_D(E)
\label{rho1dim}
,\end{equation}
which corresponds exactly to Equation (5) of Ref. \onlinecite{gasppoll93}.

If the region $\Omega$ is
connected to the outside world by tunneling barriers, so that
we have quasi-bound states in $\Omega$, the density of states
is strongly peaked for energy values corresponding to these
quasi bound states (in the limit of a closed region, the
density of states has to be a set of delta functions).
Searching for the peaks in the density of states is therefore
a possible way to find the statistics of quasi bound states.
Moreover, formula (\ref{final}) tells us that a peak in
density of states corresponds to a peak in the dwell time
for one of the incoming channels, so that, in order to
evaluate the statistics of quasi bound states \cite{wangguo94},
it is correct to search for the maxima of the dwell times
of all the possible incoming channels.

\section{Acknowledgements}
The present work has been supported by the Ministry for the University
and Scientific and Technological Research of Italy, by the Italian
National Research
Council (CNR),
and in particular by the CNR Finalized Project ``Materials
and Devices for Solid-State Electronics''.
The author gratefully acknowledges Professors C.R. Leavens
and B. Pellegrini for many stimulating
discussions and helpful comments on the manuscript, and Prof. H. Guo
for sending his manuscript prior to publication.

\appendix

\section{Two formulas for the dwell time of stationary states}

In this Appendix we want to demonstrate two formulas which relate the
expression (\ref{dwellstat}) for the dwell time to perturbative
potential approaches
\cite{sokobask87,iannpell94} and to the Green's functions for the
system under consideration.

Let us apply a perturbative potential $V$ to the region $\Omega$,
so that $V \hat{P}_{\Omega}$ is the perturbation operator to be added to
the unperturbed Hamiltonian $\hat{H}$.
The new orthonormal basis for this system is
$\{ | \phi^\pm_{n,V}(E) \rangle \}$,
where each function is a solution of the equation
\begin{equation}
( E \pm \epsilon - \hat{H} - V\hat{P}_\Omega )| \phi^{\pm}_{n,V}(E) \rangle = 0
\label{schroperturb}
,\end{equation}
for $\epsilon \rightarrow 0$ and the $n$th incoming channel.

We want to show that the dwell time
expression given by (\ref{dwellstat}) is
just the diagonal matrix element of $2 i \hbar \frac{\partial}{\partial V}$
evaluated for $V=0$, i.e.,
\begin{eqnarray}
\tau_D^{(n)}(E) & = &\left. [ 2 i \hbar \frac{\partial}{\partial V} ]_{nn}
		\right|_{V=0}
\nonumber \\
& = &
\left. \frac{ \langle \phi^\pm_{n,V}(E) | 2i\hbar \frac{\partial}{\partial V}
		| \phi^\pm_{n,V}(E) \rangle}
	    { \langle \phi^\pm_{n,V}(E) | \phi^\pm_{n,V}(E) \rangle}
		\right|_{V=0}
\label{dwellstatop}
.\end{eqnarray}
Let us point out
that both the numerator and the denominator diverge (because
the basis is Dirac-normalized).
Nevertheless their ratio is finite, as we will show.

We can write \cite{economou79}
\begin{equation}
| \phi^\pm_{n,V}(E) \rangle = | \phi_n(E)\rangle \pm V \hat{G}^\pm(E)
				\hat{P}_{\Omega}|\phi^\pm_{n,V}(E) \rangle
,
\label{phiperturb}\end{equation}
where $\hat{G}^\pm(E)$ is a solution of the equation
\begin{equation}
( E \pm \epsilon - \hat{H})\hat{G}^\pm(E) = \hat{1}
,\end{equation}
for $\epsilon \rightarrow 0$.
Substituting (\ref{phiperturb}) in (\ref{dwellstatop}) yields
\begin{equation}
\tau_D^{(n)}(E) = \pm 2i\hbar
	\frac{ \langle \phi_n(E) | \hat{G}^\pm(E) \hat{P}_{\Omega}|
						\phi_n(E) \rangle}
		{\langle \phi_n(E) |\phi_n(E) \rangle}
\label{dwellgreen}
.\end{equation}
Since in the righthand side of this equation is a diagonal matrix element,
it has to be real, therefore only the imaginary part of $\hat{G}^\pm(E)$
matters. But we know \cite{economou79} that
\begin{equation}
\mbox{Im}\{ \hat{G}^\pm(E) \} = \mp \pi
		\sum_{m=1}^N | \phi_m(E) \rangle \langle \phi_m(E) |
\end{equation}
which, substituted in (\ref{dwellgreen}) allows us to write
\begin{eqnarray}
\tau_D^{(n)}(E) & = & 2\pi\hbar
\sum_{m=1}^N \frac{
	\langle \phi_n(E) | \phi_m(E) \rangle
	\langle \phi_m(E) | \hat{P}_{\Omega} | \phi_m(E) \rangle}
	{\langle \phi_n(E) | \phi_n(E) \rangle }
\nonumber \\
& =& 2 \pi \hbar \langle \phi_n(E) | \hat{P}_{\Omega} | \phi_n(E) \rangle
\end{eqnarray}
where the last step derives from the orthogonality of the basis.
Therefore Equation (\ref{dwellstat}) is shown to be equivalent to
(\ref{dwellstatop}) and (\ref{dwellgreen}).

\end{document}